\documentclass{PoS}

\title{Bulk viscosity of strange matter and r-modes in neutron stars}

\ShortTitle{Bulk viscosity of strange matter and r-modes in neutron stars}

\author{Debarati Chatterjee\\
        Theory Division and Centre for Astroparticle Physics\\
        Saha Institute of Nuclear Physics, Kolkata-700064, India\\
        E-mail: \email{debarati.chatterjee@saha.ac.in}}

\author{\speaker{Debades Bandyopadhyay}\\
        Theory Division and Centre for Astroparticle Physics\\
        Saha Institute of Nuclear Physics, Kolkata-700064, India\\
        E-mail: \email{debades.bandyopadhyay@saha.ac.in}}

\abstract{We discuss bulk viscosity due to non-leptonic processes involving
hyperons and Bose-Einstein condensate of negatively charged kaons in neutron
stars. It is noted
that the hyperon bulk viscosity coefficient is a few order of magnitude larger
than that of the case with the condensate. Further it is found that the hyperon
bulk viscosity is suppressed in a superconducting phase. The
hyperon bulk viscosity efficiently damps the r-mode instability in neutron 
stars irrespective of whether a
superconducting phase is present or not in neutron star interior.}

\FullConference{10th Symposium on Nuclei in the Cosmos\\
                 July 27 - August 1, 2008\\
                 Mackinac Island, Michigan, USA}

\begin{document}

\section{Introduction}
Neutron stars pulsate in various modes due to its fluid perturbation.
Those oscillatory modes are classified according to different restoring 
forces. The study of Coriolis restored r-mode in neutron stars has generated 
tremendous interest in recent times \cite{Kok}. The mode becomes unstable due to
gravitational radiation. R-modes of rotating neutron stars are important 
sources of detectable gravitational waves. 
The $l=m=2$ r-mode frequency ($\omega_r$) is related to the angular frequency 
of the 
compact star as $\omega_r = \frac{2m}{l(l+1)} \Omega$ \cite{Kok}. The 
calculation of $\Omega$ depends on the equation of state (EoS) as it will be 
shown in Figure 3. Therefore, r-modes, if detected, would shed light on the EoS
and composition of matter in neutron star interior. 

Depending on the nature of strong interaction, different novel phases 
with large strangeness fraction such as, hyperon matter, Bose-Einstein 
condensate of negatively charged kaons and quark phase may appear in neutron
star interior. The r-mode instability could
be effectively suppressed by bulk viscosity due to non-leptonic processes
including hyperons \cite{Jon1,Jon2} in neutron 
star interior. 
Here we discuss bulk viscosity coefficients and the 
corresponding damping time scales due to non-leptonic weak processes involving 
hyperons and negatively charged kaons. Next we demonstrate the effect of bulk
viscosity on the damping of the gravitational radiation driven r-mode 
instability in neutron stars. Further we discuss the influence of antikaon 
condensate on the hyperon bulk viscosity. 
\section{Composition and EoS}
The knowledge about composition and EoS of dense matter is important
to investigate the oscillatory modes of compact stars. Here we consider three
different compositions for $\beta$-equilibrated and charge neutral neutron star
matter (i) including $n$, $p$, $\Lambda$ hyperons, electrons and muons, (ii) 
undergoing a first order phase transition from nuclear matter to $K^-$
condensed matter and (iii) involving a first order phase transition from 
$\Lambda$ hyperon matter to $K^-$ condensed matter.
The baryon-baryon interaction is mediated by the exchange
of $\sigma$, $\omega$ and $\rho$ mesons and two strange mesons, scalar meson,
$f_0$(975) and the vector meson
$\phi$(1020) \cite{Sch,Bog}. Similarly (anti)kaon-baryon interaction is treated
in the same footing \cite{Gle98,Gle99,Pal,Bani1,Bani2} as baryon-baryon
interaction.
\begin{figure}[h]
\centering
\includegraphics[width=6cm, height=6cm]{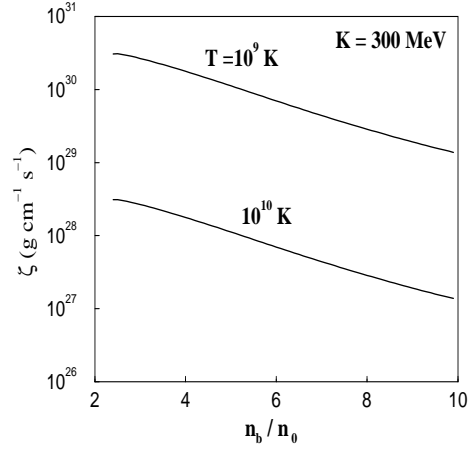}
\caption{\footnotesize{Hyperon bulk viscosity coefficient 
is plotted with normalised baryon density for
different temperatures.}}
\end{figure}
Nucleon-meson coupling constants 
are determined by reproducing nuclear matter saturation
properties such as binding energy $E/B=-16.3$ MeV, baryon density $n_0=0.153$
fm$^{-3}$, asymmetry energy coefficient $a_{\rm asy}=32.5$ MeV,
and effective nucleon masses $m^*_N/m_N = 0.78, 0.70$ corresponding two values 
of incompressibility of nuclear matter $K= 240, 300$ MeV. 
Parameters of the model are tabulated in Ref.\cite{Gle91}. Further kaon-meson
and  hyperon-meson coupling constants are obtained from Ref.
\cite{Sch,Bani2,DR3}. In this calculation, we take the value of $\Lambda$
hyperon potential depth of $-30$ MeV in nuclear matter and the value of 
antikaon optical potential depth at normal nuclear matter density 
$U_{\bar K}(n_0) = -120, -160$ MeV.

For $K=240, 300$ MeV, $\Lambda$ hyperons
appear at $2.6n_0$ and $2.3n_0$, respectively whereas $K^-$ condensation sets 
in at $3.26n_0$ for 
$K=240$ MeV and  $U_{\bar K}(n_0) = -120$ MeV and at 2.23$n_0$ for 
$K=300$ MeV and  $U_{\bar K}(n_0) = -160$ MeV.
\section{Bulk Viscosity and Critical Angular Velocity}
\begin{figure}[h]
\centering
\includegraphics[width=6cm, height=6cm]{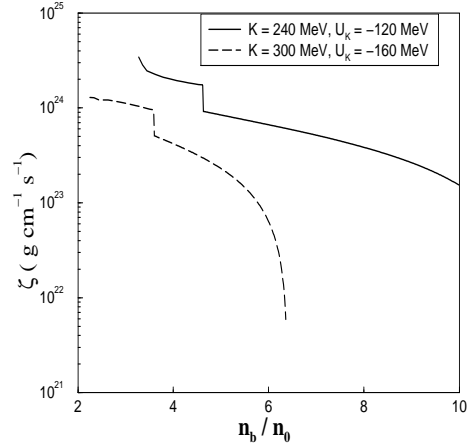}
\caption{\footnotesize{Antikaon bulk viscosity coefficient is shown as a 
function of normalised baryon density for different values of $K$ and 
$U_{\bar K}(n_0)$.}}
\end{figure}
Energy dissipation connected with pressure and density variations due to
the r-mode results in bulk viscosity. The system goes out of chemical 
equilibrium due to those variations. Weak processes bring it back to the
equilibrium. We calculate bulk viscosity due to the following two weak
processes,
\begin{eqnarray}
n + p \rightleftharpoons p + \Lambda~,\\
n \rightleftharpoons p + K^{-}~.
\end{eqnarray} 
The real part of the bulk viscosity coefficient is given by \cite{Lin02,Nar}
\begin{equation}
\zeta = \frac {P(\gamma_{\infty} - \gamma_0)\tau}{1 + {(\omega\tau)}^2}.~
\end{equation}
The relaxation times ($\tau$) for the above mentioned weak processes are taken 
from Ref.\cite{DR3,DR1,DR2}. Bulk viscosity coefficients due to weak processes 
including $\Lambda$ hyperons and the condensate as a function of normalised
baryon density are displayed in Figure 1 and Figure 2 respectively. Hyperon
bulk viscosity in Fig. 1 increases with decreasing temperature whereas the 
antikaon bulk viscosity does not vary with temperature because the condensate 
is treated at zero temperature \cite{DR3}. Further it is observed that the 
antikaon bulk viscosity coefficient is several orders of magnitude lower than 
the hyperon bulk viscosity \cite{DR3,DR1,DR2}. 

The critical angular velocity
of a neutron star is calculated by solving the overall r-mode timescale 
($\tau_r$) \cite{DR3,DR1,DR2},
\begin{equation}
{\frac {1}{\tau_r}} =  - {\frac {1}{\tau_{GR}}} + {\frac {1}{\tau_B}} +
{\frac {1}{\tau_U}} = 0~.
\end{equation}
\begin{figure}[h]
\centering
\includegraphics[width=6cm, height=6cm]{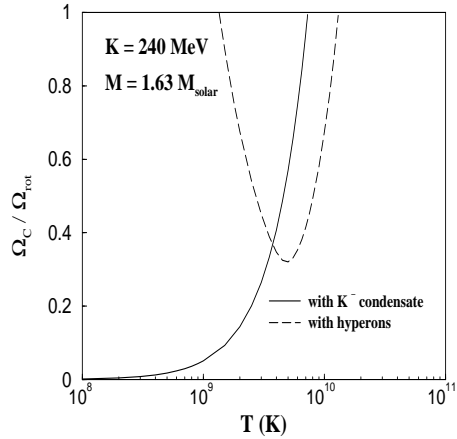}
\caption{\footnotesize{Critical angular velocity of 1.63 $M_{\odot}$ neutron 
star is shown as a function of temperature.}}
\end{figure}
Neutron stars which are rotating above critical angular velocity ($\Omega_C$), 
are unstable due to gravitational radiation whereas they are stable below 
$\Omega_C$. Critical angular velocities as a function of temperature for cases 
(i) and (ii) are exhibited in Figure 3. For case (i) denoted by the dashed 
curve, the r-mode instability of the neutron star having mass 
1.63$M_{\odot}$ is damped in the temperature regime above $10^{10}$ K because
the damping time scale ($\tau_U$) due to the modified Urca process \cite{DR1} 
involving nucleons
is comparable with that of the gravitational growth time scale ($\tau_{GR}$). 
The mode is damped by the hyperon bulk viscosity below $10^{10}$ K. 
On the other hand, the antikaon bulk viscosity can not damp the
r-mode as it is shown by the solid curve in Fig. 3.  

Next we focus on the hyperon bulk viscosity due to the non-leptonic process in
Eq. (3.1) in the presence of a $K^-$ condensate. Figure 4 shows the hyperon 
bulk viscosity coefficient as a function of normalised baryon density at a 
temperature $T = 2 \times 10^{9}$ K. The bulk viscosity in the hadronic (bold
solid line) and $K^-$ condensed (light solid line) parts of the mixed phase
and the pure condensed phase (dashed line) are shown in the figure. The hyperon
bulk viscosity in the antikaon condensed matter is suppressed than that of the
hadronic phase. It is observed that the hyperon bulk viscosity in $K^-$
condensed matter might damp the r-mode instability effectively \cite{DR4}. 
\section{Summary}
We have investigated bulk viscosity due to non-leptonic processes involving 
$\Lambda$ hyperons and $K^-$ condensate in neutron stars. It is noted that the
hyperon bulk viscosity coefficient is a few orders of magnitude higher than the
antikaon bulk viscosity coefficient. Hyperon bulk viscosity efficiently damps 
the r-mode instability. 

\begin{figure}[t]
\begin{center}
\includegraphics[width=6cm,height=6cm]{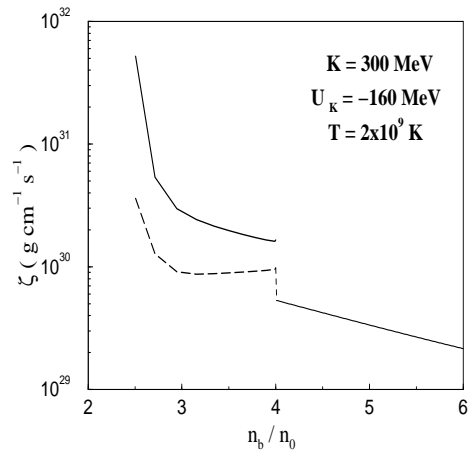}
\caption{Hyperon bulk viscosity coefficient in $K^-$ condensed matter is shown 
as a function of normalised baryon density.}
\end{center}
\end{figure}

\end{document}